\documentclass[journal]{IEEEtran}

\usepackage[T1]{fontenc}
\usepackage{textcomp}
\usepackage{amsmath,amssymb,amsfonts}
\usepackage{graphicx}
\usepackage{xspace}
\usepackage{url}
\usepackage{makecell}
\usepackage{array}
\usepackage{acronym}
\usepackage{adjustbox}
\usepackage{float}
\usepackage{latexsym}
\usepackage{courier}
\usepackage[table,xcdraw]{xcolor}
\usepackage{diagbox}
\usepackage{comment}
\usepackage{longtable}
\usepackage{tabularx}
\usepackage{cancel}
\usepackage{array}
\usepackage{multirow}
\usepackage{booktabs}
\usepackage{listings}
\usepackage{caption}
\usepackage{subfig}
\usepackage{tikz}
\usepackage{fmtcount}
\usepackage{algorithm}
\usepackage{algpseudocode}
\usepackage{tcolorbox}
\usepackage{hyperref}

\usepackage{tikz}
\usetikzlibrary{positioning,arrows.meta}
\usepackage{pifont}
\definecolor{dkgreen}{rgb}{0,0.6,0}
\definecolor{gray}{rgb}{0.5,0.5,0.5}
\definecolor{mauve}{rgb}{0.58,0,0.82}

\tcbset{
  colback=gray!5!white,
  colframe=gray!40!black,
  boxrule=0.5pt,
  arc=2pt,
  left=4pt, right=4pt, top=4pt, bottom=4pt
}
\newenvironment{summarybox}{\begin{tcolorbox}\itshape}{\end{tcolorbox}}
\newcommand{\sysname}{\textsf{TaintRadar}\xspace}

\lstdefinestyle{java}{
  language=Java,
  aboveskip=1mm,
  belowskip=1mm,
  showstringspaces=false,
  columns=flexible,
  basicstyle={\small\ttfamily},
  numbers=left,
  numberstyle=\tiny\color{gray},
  numbersep=5pt,
  keywordstyle=\color{blue},
  commentstyle=\color{dkgreen},
  stringstyle=\color{mauve},
  breaklines=true,
  breakatwhitespace=true,
  tabsize=1,
  xleftmargin=0pt,
  xrightmargin=0pt,
  framexleftmargin=0pt,
  framexrightmargin=0pt
}

\newtheorem{definition}{Definition}

\def\BibTeX{{\rm B\kern-.05em{\sc i\kern-.025em b}\kern-.08em
    T\kern-.1667em\lower.7ex\hbox{E}\kern-.125emX}}

\begin{document}

\title{\sysname{}: Semantic-Aware Taint-Style Vulnerability Detection via Augmented Code Property Graphs}

\author{
\IEEEauthorblockN{
Elie Rizk\IEEEauthorrefmark{1},
Firas Ben Hmida\IEEEauthorrefmark{2},
Birhanu Eshete\IEEEauthorrefmark{3}\\
}
\IEEEauthorblockA{
\IEEEauthorrefmark{1}elirizk@umich.edu,\;
\IEEEauthorrefmark{2}fbhmida@umich.edu,\;
\IEEEauthorrefmark{3}birhanu@umich.edu\\
Department of Computer and Information Science\\
University of Michigan-Dearborn, Michigan, USA
}
}

\title{\sysname{}: Semantic-Aware Taint-Style Vulnerability Detection via Augmented Code Property Graphs}

\maketitle

\begin{abstract}
Despite significant advances, static vulnerability analysis suffers from three critical limitations: coarse sanitization modeling (treating validation as a binary barrier), database blindness (breaking taint tracking across persistence layers), and shallow object-oriented analysis (missing field-level and interprocedural data flows). These flaws stem from a common root cause: Code Property Graph (CPG)-based taint analyses lack a compositional semantic layer to jointly model sanitization, persistence, and object aliasing. Consequently, existing tools generate excessive false positives or miss critical attack paths entirely.

To address these limitations, we present \sysname{}, an approach that systematically augments CPGs with three independent semantic analysis layers. First, vulnerability-typed sanitization computes node-level safety guarantees using transfer functions and context-sensitive parameter binding. Second, persistence-aware propagation integrates database schema constraints and query safety analysis to track multi-script attack paths traversing shared database states. Finally, object-aware reaching definitions combine calling and bounded variable alias contexts to precisely model object-field mutations across method boundaries.

We evaluate \sysname{} on both synthetic benchmarks and real-world systems. On the SARD benchmark, \sysname{} drastically reduces false positives while maintaining 83\% overall accuracy. Deployed across 19 real-world PHP applications, it rediscovered the majority of known CVEs and uncovered 30 confirmed zero-day vulnerabilities (27  SQL injection and 3 stored XSS) that have already been assigned CVE identifiers. These results demonstrate that semantic-aware graph augmentation significantly improves the precision, coverage, and practical utility of static taint analysis.
\end{abstract}


\section{Introduction}\label{sec: intro} 
Taint-style vulnerabilities represent a critical and highly prevalent class of software security weaknesses in which untrusted or malicious input propagates unchecked from untrusted external inputs (\emph{sources}) to execution-critical operations (\emph{sinks}), such as dynamic command execution or database interpreters. These vulnerabilities are particularly rampant in PHP-based web applications, which natively process massive volumes of user-controlled inputs through dynamically typed, string-heavy operations and database-backed workflows.

Despite significant progress in static analysis frameworks leveraging taint tracking and Code Property Graphs (CPG), existing techniques~\cite{KAVE,Yama,10.1145/3442381,10.1145/1250734,10.5555/2028067,10.1145/204670,10.1145/354860,1624016, 10.1145/2660267.2660363, 10.5555/2671225.2671288, DBLP:journals/jksucis/MarashdihZS23} face severe architectural limitations. Many prior solutions remain narrow in scope, relying on coarse syntactic pattern matching or naively tracing source-to-sink paths without robust, context-aware verification, which triggers prohibitive false positive rates. Current CPG-based vulnerability detection frameworks lack the ability to model or enforce the semantic operations governing how data is transformed between decoupled program layers. Object-oriented features and persistent database states are frequently abstracted away or completely ignored~\cite{haq2024lucidframeworkreducingfalse}, breaking data flow continuity. While hybrid approaches attempt to address these gaps via dynamic execution engines~\cite{AlhuzaliNAVEX18} or localized call-graph refinements~\cite{app13169240}, they fail to capture the deep semantics of object-field interactions and cross-script database persistence, leaving a massive attack surface completely unanalyzed. These deficiencies underscore a fundamental question in static vulnerability analysis: \textbf{\emph{how can CPG-based taint analysis be made semantic-aware rather than purely structural, while remaining scalable to large, real-world applications?}}

In response to this question, we present \sysname{}, a system that treats taint-style vulnerability discovery as a sequence of independent, composable graph augmentation operators executed over an optimized base CPG representation. Rather than reinventing the underlying parsing architecture, \sysname{} layers domain-specific security semantics directly onto the graph representation. First, target application code is lifted into a base CPG that unifies abstract syntax trees (AST), control flow graphs (CFG), and program dependence graphs (PDG). Atop this foundation, \sysname{} deploys a language-specific knowledge layer that explicitly maps vulnerability-relevant characteristics—ranging from superglobal request inputs to security-critical execution sinks—grounding the downstream analysis engine in explicit program semantics rather than fragile ad hoc heuristics.

The analysis then proceeds through three independent graph augmentation stages. A sanitization inference layer computes node-level lattice guarantees for individual vulnerability classes through targeted forward data-flow reasoning, making explicit exactly where and how validation logic applies. Simultaneously, \sysname{} evaluates persistence-aware taint propagation by executing column-specific database schema constraints and query safety analyses; this allows the engine to seamlessly track data flows that traverse application-database boundaries across otherwise disconnected scripts. To prevent structural noise, an object-aware reaching-definition layer enriches the graph by concurrently resolving calling contexts and bounded object-field mutations across method boundaries. Once these semantic layers are woven into the graph, a unified backward reachability traversal traces precise paths from unsanitized sinks to untrusted sources, yielding a high-confidence set of actionable vulnerabilities. As an optional validation step, \sysname{} cross-references its findings with public CVE records to automatically verify coverage against documented exploits.

Our comprehensive evaluation demonstrates that \sysname{} significantly outperforms state-of-the-art static analysis tools in detecting complex taint-style vulnerabilities within real-world PHP applications. On the synthetic SARD~\cite{nist2023sard} benchmark, \sysname{} drastically curtails false positive rates while maintaining a robust 83\%  overall detection accuracy and an F1-score of 76\% on highly complex, structurally diverse test cases. In evaluations across popular open-source software, \sysname{} successfully rediscovered flaws matching 83.33\% of known historical CVEs. More importantly, \sysname{} uncovered 30 previously unknown zero-day vulnerabilities that have since been validated and assigned public CVE IDs, including 27  SQL injection and 3 stored cross-site scripting (XSS) vulnerabilities across 9 applications.  These results demonstrate that semantic-aware graph augmentation provides a scalable, highly precise blueprint for modern static analysis.

This paper makes the following key contributions:
\begin{itemize}
    \item \textbf{Semantic-Aware Sanitization Analysis}: We introduce a sanitization-aware data-flow framework that computes precise, node-level lattice guarantees using formal transfer functions and context-sensitive interprocedural parameter binding, significantly eliminating false positives compared to traditional binary taint barriers.

    \item \textbf{Database-Aware Taint Analysis}: We systematically integrate database schema constraints and column-specific query safety analyses into static taint analysis. By drawing targeted persistence-aware dependency edges across write/read boundaries, \sysname{} uncovers cross-script attack vectors—such as  SQL injections and stored XSS—that completely break data flow continuity in conventional CPG models.
    
    \item \textbf{Object Dependency Graph Integration}: We enhance CPG capabilities with Object Dependency Graphs (ODGs) derived from a context-sensitive reaching definition analysis. By introducing explicit $k$-bounded calling and variable stacks, the engine precisely tracks field mutations across constructors and object boundaries without risking state-space explosion or infinite loops.
    
    \item \textbf{Comprehensive Evaluation and Open-Source Implementation}: We provide an extensive empirical evaluation across controlled benchmarks and 19 production-grade applications. We demonstrate decisive performance gains over SOTA tools and release our complete implementation to the research community at:\\ {\small \url{https://github.com/um-dsp/TaintRadar/tree/artifacts_code}}.
\end{itemize}
\section{Background on  Code Property Graphs}\label{sec: bground}

Yamaguchi et al. \cite{YamaguchiCPG14} first introduced CPGs to model and discover vulnerabilities through static code analysis. A CPG is a representation of the source code combining multiple classical program analysis abstractions in the form of Abstract Syntax Trees (AST), Control Flow Graphs (CFG), and Program Dependency Graphs (PDG). As a result of merging the structure, control flow, and data dependencies of the source code, researchers and practitioners can frame vulnerability analysis as graph traversals on the CPG. 

Given an application's CPG, most vulnerabilities in the source code can be modeled as information flow problems that violate confidentiality or integrity. For example, a Cross-Site Scripting (XSS) attack involves an attacker-controlled input reaching a sensitive sink displaying it to a console, browser, database, or API response. An SQL injection involves a path from an attacker-controlled source to an SQL query execution, allowing unrestricted escaping or modification of the executed statement. Consequently, Yamaguchi et al. \cite{YamaguchiCPG14} grouped CPG vulnerability descriptions in two broad categories. \textbf{Syntax-only vulnerability descriptions} rely solely on the code's AST to detect vulnerable code, such as matching all non-constant arguments to the {\tt print} function. This type of description fails to capture attacker control or the relationship between different statements, ultimately resulting in missed coverage or multiple false positives. Accordingly, \textbf{taint-style vulnerability descriptions} are represented by syntax-only descriptions of attacker-controlled sources, security sensitive sinks and sanitizers. A path matches a taint-style description if a data dependency path exists from source to sink without passing by sanitizer nodes.

Backes et al. \cite{BackesPHP17} laid the foundational work of applying CPG traversal to the high-level dynamic scripting language that is PHP. Their research defined attacker-controlled input and listed different vulnerability classes relevant for PHP static code analysis. For taint-style vulnerabilities, potential sources include all parameters that can be transferred through an HTTP request, as they constitute attacker-controlled inputs. PHP wraps HTTP request parameters in various associative arrays including:
\verb|$_GET|, \verb|$_POST|, \verb|$_COOKIE|, \verb|$_REQUEST|, \verb|$_SERVER|, and \verb|$_FILES|. As for the sink nodes, the authors include all sensitive functions for each vulnerability, e.g. \verb |mysql_query| for an SQL injection attack or \verb|echo| for XSS.

\section{Challenges \& Running Examples}\label{sec:challenge}
We use the running examples in Listings~\ref{lst:input}, \ref{lst:display}, and \ref{lst:oo_input} to motivate the core limitations of standard Code Property Graph (CPG) implementations. 

Listing~\ref{lst:input} shows a classic PHP example of attacker-controlled input originating from browser cookies (\texttt{\$user\_id}) and HTTP POST requests (\texttt{\$comment\_title} and \texttt{\$comment}). While some variables are sanitized against Cross-Site Scripting (XSS)—either explicitly via functional sanitizers or implicitly through strict type coercion—an unsanitized value directly reaches the SQL statement construction. 

The values stored in the database are later retrieved by the separate script in Listing~\ref{lst:display} and rendered to the output buffer. The first code sample is vulnerable to SQL Injection (SQLi) attacks through both the \texttt{\$comment\_title} and \texttt{\$comment} parameters, while the second is vulnerable to a Stored (Persistent) XSS attack through the \texttt{\$comment} field. 

This scenario underscores the need for a principled static analysis tool that computes precise, node-level sanitization guarantees per vulnerability type and tracks vulnerable data flows that span persistent storage boundaries such as a database.

\begin{table}[!ht]
\centering
\caption{Database Schema Mapping}
\label{tab:db_schema}
\smallskip
\resizebox{0.48\textwidth}{!}{%
\begin{tabular}{|l|l|l|l|} \hline\hline
\textbf{Table} & \textbf{Column} & \textbf{Type} & \textbf{Safe Type} \\ \hline
blog\_posts & user\_id & TINYINT & True \\ \hline
blog\_posts & comment\_title & VARCHAR(20) & False \\ \hline
blog\_posts & comment & VARCHAR(100) & False \\ \hline\hline
\end{tabular}%
}
\end{table}

\begin{lstlisting}[language=PHP, caption={Getting user input (input.php)}, label={lst:input}]
<?php
$user_id = (int)$_COOKIE["uid"];
$comment_title = htmlentities($_POST['comment_title']);
$comment = $_POST['comment'];

if (isset($user_id) && isset($comment_title) && isset($comment)) {
    $link = mysqli_connect("localhost", "user", "psswd", "db");
    $sql = "INSERT INTO blog_posts VALUES (" . $user_id . ", '" . $comment_title . "', '" . $comment . "')";
    mysqli_query($link, $sql);
}
?>
\end{lstlisting}

\begin{lstlisting}[language=PHP, caption={Displaying data (display.php)}, label={lst:display}]
<?php
$link = mysqli_connect("host", "user", "psswd", "db");
$sql = "SELECT * FROM blog_posts";
$result = mysqli_query($link, $sql);
$rows = mysqli_fetch_all($result, MYSQLI_ASSOC);

foreach ($rows as $row) {
    echo "User: " . $row["user_id"] . " - Comment: " . $row["comment_title"] . ":\n" . $row["comment"];
}
?>
\end{lstlisting}

\begin{lstlisting}[language=PHP, caption={PHP Object-oriented running example (PostDAO.php)}, label={lst:oo_input}]
<?php
class PostDAO {
    public $sql;

    public function __construct($uid, $title, $comment) {
        $safe_uid = (int)$uid;
        $safe_title = htmlentities($title);     
        $this->sql = "INSERT INTO blog_posts VALUES (" . $safe_uid . ", '" . $safe_title . "', '" . $comment . "')";
    }

    public function addTenantFilter($tenantId) {
        $this->sql .= " AND tenant_id=" . (int)$tenantId;
    }

    public function exec($link) {
        mysqli_query($link, $this->sql);
    }
}

$dao = new PostDAO($_COOKIE["uid"], $_POST["comment_title"], $_POST["comment"]);
$dao->addTenantFilter($_GET["t"]);
$dao->exec(mysqli_connect("localhost", "user", "psswd", "db"));
?>
\end{lstlisting}

This work aims to address three fundamental challenges exposed by these scenarios:

\paragraph{Challenge 1: Vulnerability-Typed Sanitization Filtering} \label{chall1}
Existing static analysis tools fail to provide precise, vulnerability-specific sanitization guarantees, leading to high false-positive rates when sanitization is conditional, partial, or context-dependent. A security framework must augment the base CPG with semantic-aware contexts to compute accurate sanitization labels at the node level per vulnerability type. 

In Listing~\ref{lst:input}, the \texttt{\$user\_id} variable is implicitly sanitized against injection through strict type coercion to an integer, and \texttt{\$comment\_title} is properly sanitized against XSS using \texttt{htmlentities}, yet it remains completely vulnerable to SQL injection. Conversely, \texttt{\$comment} remains entirely un-sanitized. Implicit type conversions can also occur at the database layout layer where \texttt{user\_id} is securely kept as a numeric type, rendering any stored XSS payloads ineffective within that field. 

The core challenge lies in computing precise, multi-vulnerability sanitization signatures within the graph while dynamically parsing varying language features such as PHP's loose type juggling.

\paragraph{Challenge 2: Database-Aware Cross-Boundary Vulnerability Detection} \label{chall2}
Traditional static analysis tools evaluate application code in isolation from database structures, completely missing vulnerability paths that traverse persistent storage boundaries. In our running example, the taint chain initiates during the database insertion step in \texttt{input.php} (Listing~\ref{lst:input}) and completes via unsafe retrieval and rendering inside \texttt{display.php} (Listing~\ref{lst:display}), despite the lack of direct static code connections (such as file inclusions or shared function calls) between the scripts. 

Resolving this requires parsing database configurations to verify column constraints (per Table~\ref{tab:db_schema}), auditing query structural safety, and mapping attack vectors across persistence barriers. The core challenge involves modeling how tainted parameters committed to storage in one request lifecycle propagate into separate execution scopes, particularly under differing sanitization requirements for data storage versus subsequent output generation.

\paragraph{Challenge 3: High-Fidelity Tracking of Object-Field Data Flows} \label{chall3}
Standard CPG representations often fail to capture fine-grained data dependencies for individual object field attributes. This introduces severe visibility gaps in modern object-oriented codebases where attacker-controlled variables mutate shared internal object states across method and constructor boundaries. 

In Listing~\ref{lst:oo_input}, external user values are combined into the class property \texttt{\$this->sql} inside the constructor, conditionally appended via a secondary method invocation (\texttt{addTenantFilter()}), and ultimately executed inside \texttt{exec()}. Although \texttt{\$user\_id} undergoes implicit integer casting and \texttt{\$comment\_title} is safely filtered against XSS, the un-sanitized string \texttt{\$comment} flows directly into the database engine via object field manipulation. 

The core challenge lies in extending the CPG infrastructure with object-field dependency tracking that models field read/write assignments, reference aliasing, interprocedural parameter bindings, and class framework patterns while preserving rigorous context-sensitivity.
\section{Approach}\label{sec: approach}

\begin{figure*}[t!]
    \centering
    \includegraphics[scale=0.55]{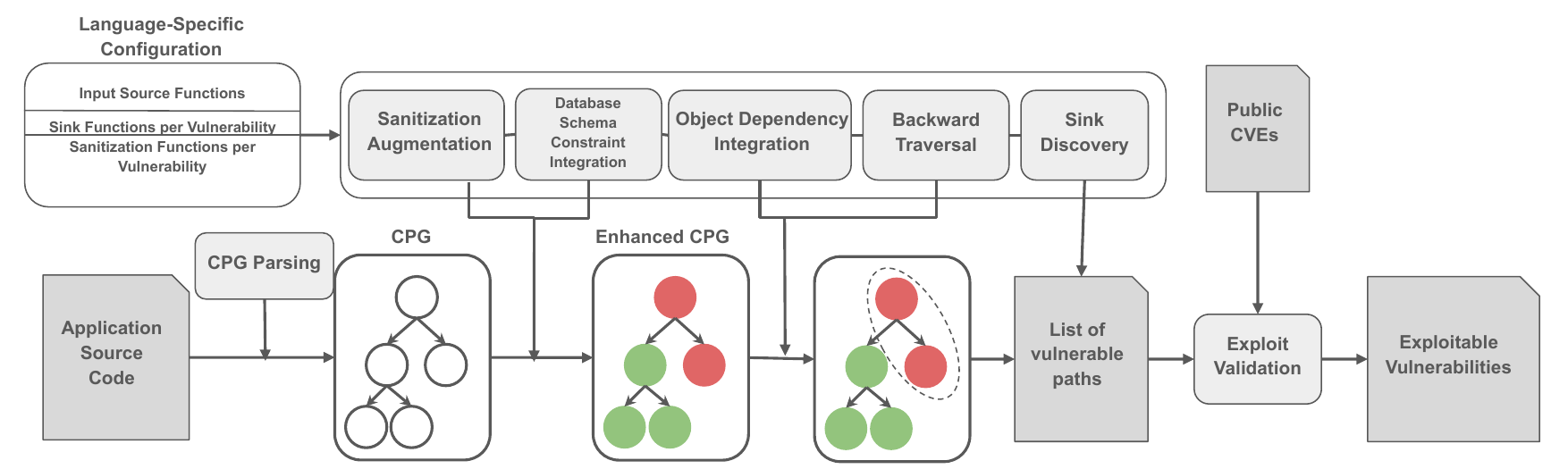}
    \caption{\sysname{} system overview.}
    \label{fig:overview}
\end{figure*}
\subsection{\sysname{} Design Overview}

\sysname{} is built on a central guiding principle: vulnerability discovery can be reduced to a reachability problem over a semantically enriched CPG. Instead of introducing a new analysis formalism, \sysname{} incrementally augments a base CPG with critical security-relevant semantic relations. As illustrated in Figure~\ref{fig:overview}, the framework ingests source code to construct a base CPG, then sequentially applies three independent, composable augmentation operators designed to systematically bridge historical precision and recall gaps:

\begin{itemize}
    \item \textbf{Sanitization Augmentation} computes vulnerability-specific, node-level safety guarantees via context-sensitive parameter binding, avoiding the coarse binary barrier modeling of prior tools.
    \item \textbf{Database Integration} establishes persistence-aware dependencies by leveraging schema constraints and query safety properties, explicitly linking related write and read operations across distinct scripts.
    \item \textbf{Object Dependency Augmentation} injects precise object-field dependencies derived from context-sensitive reaching definitions to reliably track field mutations across method boundaries.
\end{itemize}

Following these augmentations, \sysname{} executes a unified \textbf{Backward Traversal} over the enriched dependency graph to discover source-to-sink paths where an attacker-controlled source can reach a sensitive sink without a sufficient sanitization guarantee. The resulting candidate exploit paths undergo automated \textbf{Exploit Validation} by cross-referencing public CVE databases to flag known vulnerabilities. Unmatched findings are then manually audited to verify exploitability and identify zero-days. As detailed in \S\ref{sec: eval}, this modular architecture significantly improves detection coverage and precision while scaling efficiently to large, real-world production codebases.
\subsection{Preliminaries and Definitions}
To ease our explanation of \sysname{} design details, we first introduce preliminaries and definitions.

Let $G = (V, E, \lambda, \mu)$ be a Code Property Graph, where
{\small
\[
\begin{array}{rcl}
    V & : & \text{set of AST nodes}\\[4pt]
    E = E_{AST} \cup E_{CFG} \cup E_{PDG} & : & \text{set of AST, CFG, and PDG edges}\\[4pt]
    \lambda : E \to L & : & \text{edge labeling function}\\[4pt]
    \mu : (V \cup E) \times K \to S & : & \text{node and edge property function}
\end{array}
\]
}
The edge labeling function $\lambda$ assigns a label $L$ representing the origin of the edge (whether it’s an AST, CFG or DDG edge). The property function $\mu$ assigns a property $\mu(v, k)=S$ for key $K$ simplified as $v_k$, e.g. a boolean value representing sanitization status of a node.

\begin{definition}[MATCH Traversal]
Following the original CPG framework~\cite{YamaguchiCPG14}, we define the reusable MATCH traversal for node filtering:
\[
\text{MATCH}_p(V) = \text{FILTER}_p \circ \text{TNODES}(V)
\]
where
{\small
\[
\begin{aligned}
    \text{TNODES}(V) & : \; \text{traverses from roots in $V$ to all AST nodes} \\[4pt]
    \text{FILTER}_p  & : \; \text{filters nodes according to predicate $p$} \\[4pt]
    p : V \to \{\text{true}, \text{false}\} & : \; \text{node property predicate}
\end{aligned}
\]

}
\end{definition}

\begin{definition}[Vulnerability Configuration]\label{def:vuln-config}
A vulnerability is characterized by its sanitization and sink functions
{\small
\[
\mathcal{V} = (\Phi, \Sigma)
\]
where
\[
\begin{aligned}
    \Phi &\subseteq \mathcal{F} && \text{is the set of sanitization functions}\\
    \Sigma &\subseteq \mathcal{F} && \text{is the set of sink functions}\\
    \mathcal{F} &&& \text{is the universe of all functions}.
\end{aligned}
\]
}
\end{definition}

\begin{definition}[Global Attack Sources]\label{def:source-def}
Attack sources are vulnerability-independent and capture (i) syntactic identifiers that directly reference attacker-controlled inputs and (ii) API calls that return attacker-controlled values. They are defined as
$\Omega = \Omega_{\text{id}} \cup \Omega_{\text{func}}$, where
$\Omega_{\text{id}} \subseteq \mathcal{S}$ is the set of untrusted source identifiers, and $\Omega_{\text{func}} \subseteq \mathcal{F}$ is the set of untrusted source functions. Intuitively, $\Omega_{\text{id}}$ captures sources such as \texttt{\$\_GET} or \texttt{\$\_POST} in PHP.
\end{definition}

\begin{definition}[Database Schema]\label{def:db-definition}
Let $\mathcal{D} = (T, C, \tau)$ be a database schema where
{\small
\[
\begin{array}{rcl}
    T & : & \text{set of table names}\\[4pt]
    C & : & \text{set of column names}\\[4pt]
    \tau: T \times C \to \{0,1\} & : & \text{column safety function}
\end{array}
\]
}
where $\tau(t, c) = 1$ indicates column $c$ in table $t$ has a \emph{safe type} such as INT or DATETIME and $\tau(t, c) = 0$ indicates an \emph{unsafe type} such as VARCHAR or TEXT.
\end{definition}

\subsection{\sysname{} Design Details}

\textbf{Language-Specific Configuration}.
Since an application's CPG is an abstract representation of the code, \sysname{}’s high level analysis isn't tied to the particular programming language of the application. However, a starting a language-specific configuration is required to identify relevant nodes such as input, sinks or sanitizers.
For PHP codebase, the configuration defines:
\begin{itemize}
    \item Attacker-controlled sources cover every entry point through which request data reaches the application: the superglobal arrays (e.g., \texttt{\$\_GET}, \texttt{\$\_POST}, \texttt{\$\_COOKIE}), the \texttt{filter\_input} API with its \texttt{INPUT\_*} channel constants, and header-retrieval functions such as \texttt{apache\_request\_headers()} and \texttt{getallheaders()}.
    \item Vulnerability-specific sanitization functions ~\cite{AlhuzaliNAVEX18} such as \texttt{mysql\_escape\_string} to protect against SQL injections or \texttt{htmlentities} for XSS.
    \item Globally safe functions and language constructs identified by parsing the signatures of PHP functions from \href{https://github.com/jetbrains/phpstorm-stubs}{JetBrains' PHP stubs} and flagging any function that returns safe types or a hash (such as \texttt{isset}, \texttt{intval}, \texttt{mhash}), i.e., functions whose output is guaranteed to belong to a type that cannot carry injectable syntax, regardless of the input.
    \item Validation constructs ($\Psi$) including validating predicates      
    (e.g., \texttt{is\_numeric}, \texttt{ctype\_digit}), type-reporting functions
    (e.g., \texttt{gettype}), pattern-matching functions (e.g., \texttt{preg\_match}),
    and the \texttt{filter\_*} family with the subset $F_{\text{safe}}$ of filter
    flags that reduce a value to a safe type.
    \item Security-sensitive sinks such as \texttt{mysql\_query} for SQL injection or \texttt{echo} for XSS. The vulnerabilities covered include SQL injection, command execution, code injection, file inclusion, Cross-Site Scripting (XSS), arbitrary file reads or writes, and session fixation attacks.
\end{itemize}
The above PHP-specific schema can be viewed and generated from our artifacts. It isn't fine-tuned on any particular application, rather it reflects the language’s built-in security critical constructs. This specification phase directly supports our sanitization and sink analysis, grounding the rest of the pipeline in precise, context-aware logic.

\textbf{Sanitization Augmentation}.
\sysname{} relies on our Sanitization Analysis algorithm: a forward data flow analysis that processes the CPG to compute sanitization guarantees per vulnerability type, effectively addressing Challenge 1 in Section~\ref{sec:challenge}. We define our domain and function below.

\begin{definition}[Sanitization Domain]
The analysis operates over the two-point lattice $(\{0, 1\}, \leq)$, where
\renewcommand{\arraystretch}{1.0}
\[
\begin{array}{rcl}
   0 & : & \textbf{unsanitized} \text{ (potentially tainted)} \\
   1 & : & \textbf{sanitized} \text{ (guaranteed safe)}\\
   0 \leq 1 & : & \text{``sanitized'' is better than ``unsanitized''}
\end{array}
\]
\end{definition}

\paragraph{Sanitization Function}
For vulnerability type $\mathcal{V} = \hspace{-0.1cm} (\Phi, \Sigma)$, we compute the node-level sanitization function
\[
\mathcal{S}^{\mathcal{V}}(v) =
\mathcal{T}^{\mathcal{V}}\!\left(
   v, \{\mathcal{S}^{\mathcal{V}}(u) : u \in \text{PRODUCERS}(v)\}, \pi^{\mathcal{V}}
\right),
\]
where
\[
\renewcommand{\arraystretch}{1.0}
\begin{array}{rcl}
   \text{PRODUCERS}(v) &=& \hspace{-0.3cm} \{u \in V : (u,v) \in E \land \lambda(u,v) = DDG\}, \\[-2pt]
   && \hspace{-0.3cm} \text{data-dependency predecessors of $v$}, \\[4pt]
   \mathcal{T}^{\mathcal{V}} &:& V \times \mathcal{P}(\{0,1\}) \times \Pi \to \{0,1\}, \\[-2pt]
   && \text{transfer function}, \\[4pt]
   \pi^{\mathcal{V}} &\in& \Pi, \quad \text{parameter context for}\\ && \text{interprocedural analysis}.
\end{array}
\]

The sanitization function applies different logic based on the node type. We define the transfer function $\mathcal{T}^{\mathcal{V}}$ for each node category as follows.

\textbf{Literal values} are considered sanitized by default:
\begin{align}
&\text{MATCH}_\text{literal}(v) \Rightarrow \mathcal{T}^{\mathcal{V}}(v, \_, \_) = 1
\end{align}

\textbf{Field identifiers} are handled by checking against known tainted identifiers, global constants (such as magic constants in PHP), and a pre-built constant table (a mapping between defined constants and their values):
\begin{align}
&\text{MATCH}_{\text{field}}(v) \Rightarrow \nonumber \\
&\mathcal{T}^{\mathcal{V}}(v, D, \pi^{\mathcal{V}}) =
\begin{cases}
0 & v_{\text{name}} \in \Omega_{\text{id}} \\
1 & v_{\text{name}} \in \Xi_{\text{magic}} \\
\mathcal{S}^{\mathcal{V}}(\text{ConstTable}[v_{\text{name}}]) &
  \begin{array}[c]{@{}l@{}} v_{\text{name}} \\ \in \text{ConstTable} \end{array} \\
\bigwedge_{d \in D} d & \text{otherwise}
\end{cases} \label{eq:field}
\end{align}

For \textbf{identifiers}, we traverse their reaching definition edges to check for
prior sanitization, type casting to a safe type, pass by reference to a function call, and validation guards:
\begin{align}
&\text{MATCH}_{\text{identifier}}(v) \Rightarrow \nonumber \\
&\mathcal{T}^{\mathcal{V}}(v, D, \pi^{\mathcal{V}}) = \nonumber \begin{cases}
0 & v_{\text{name}} \in \Omega_{\text{id}} \lor v_{\text{type}} \in \Omega_{\text{type}} \\
1 & v_{\text{type}} \in \Sigma_{\text{safe}} \\
1 & \text{GUARDED}(v) \\
\mathcal{T}^{\mathcal{V}}_{\text{byref}}(v, \pi^{\mathcal{V}}) & v \text{ passed by reference} \\
\bigwedge_{d \in D} d & \text{otherwise}
\end{cases}
\end{align}

\paragraph{Validation Guards}
PHP code often validates a value in a branch condition rather than
transforming it such as \texttt{if (is\_numeric(\$x))}. We treat a use
of $x$ as \emph{guarded} if it lies on a branch of an enclosing
\texttt{if}, \texttt{while} or \texttt{for} statement whose condition
must validate $x$ for that branch to execute, and $x$ is not reassigned
or retyped within the branch. A condition validates $x$ if it
applies a predicate in $\Psi$ to $x$, compares a type-reporting call
on $x$ to a safe type (e.g., \texttt{gettype(\$x) == "integer"}), or
matches $x$ against an anchored whitelist pattern
(e.g., \texttt{preg\_match("/\textasciicircum[0-9]*\$/", \$x)}).

For identifiers passed by reference, we follow their corresponding parameter within the function body:
\begin{align}
&\mathcal{T}^{\mathcal{V}}_{\text{byref}}(v, \pi^{\mathcal{V}}) = \nonumber \\
&\begin{cases}
1 & \begin{array}[c]{@{}l@{}} v \in \text{ByRefArgs}(\texttt{settype}) \\ \land\ \text{ARG}^2 \in \Sigma_{\text{safe}} \end{array} \\
\mathcal{S}^{\mathcal{V}}(\text{ReturnDef}(f, i_v)) & v \in \text{ByRefArgs}(f) \\
\bigwedge_{d \in \text{PRODUCERS}(v)} \mathcal{S}^{\mathcal{V}}(d) & \text{otherwise}
\end{cases}
\end{align}
where $\text{ByRefArgs}(f)$ denotes arguments passed by reference to
function $f$, $i_v$ is the parameter index of argument $v$, and
$\text{ReturnDef}(f, i)$ returns the return-point definition of the
$i^{th}$ parameter of $f$.

\textbf{Function calls} fall into two categories: user-defined or undefined. For calls to undefined methods (no call body present in the application code, including built-in operators like \texttt{=} and \texttt{+}), we apply the following heuristics: known sanitization functions, functions that return safe types (e.g., hashing, boolean operators, casting functions), and safe \texttt{filter\_*} functions are treated as sanitized, otherwise functions are considered sanitized only if all their arguments are sanitized. For user-defined methods, we implement an interprocedural static analysis.
\begin{align}
&\text{MATCH}_{\text{call}}(v) \Rightarrow
  \mathcal{T}^{\mathcal{V}}(v, D, \pi^{\mathcal{V}}) = \nonumber \\
&\begin{cases}
1 & v_{\text{name}} \in \Phi \\
1 & v_{\text{type}} \in \Sigma_{\text{safe}} \\
1 & v_{\text{name}} \in \Psi_{\text{filter}} \land \text{flags}(v) \subseteq F_{\text{safe}} \\
0 & v_{\text{name}} \in \Omega_{\text{func}} \lor v_{\text{name}} \in \Psi_{\text{filter}} \\
\mathcal{T}_{\text{interproc}}^{\mathcal{V}}(v, \pi^{\mathcal{V}}) & \text{user-defined} \\
\bigwedge_{d \in D} d & \text{otherwise}
\end{cases}
\end{align}
where $\text{flags}(v)$ are the \texttt{FILTER\_*} constants passed to $v$, so
\texttt{FILTER\_SANITIZE\_NUMBER\_INT} sanitizes but \texttt{FILTER\_UNSAFE\_RAW}
does not.

\paragraph{Interprocedural Analysis}
For user-defined function calls, we perform context-sensitive analysis with parameter binding.

$$
\mathcal{T}_{\text{interproc}}^{\mathcal{V}}(v, \pi^{\mathcal{V}}) = \mathcal{S}^{\mathcal{V}}(\text{ReturnBlock}(\text{Callee}(v)))
$$

where $\text{ReturnBlock}$ corresponds to all CPG nodes returned by the method to the caller, and the parameter context is defined as:

$$
\pi^{\mathcal{V}}_v = [\mathcal{S}^{\mathcal{V}}(\text{ARG}^1_v), \mathcal{S}^{\mathcal{V}}(\text{ARG}^2_v), \ldots, \mathcal{S}^{\mathcal{V}}(\text{ARG}^n_v)]
$$

\paragraph{Context Binding}
For a function call $v$ with callee $f$ and arguments $\{\text{ARG}^i_v\}_{i=1}^n$, the parameter context $\pi^{\mathcal{V}}_v$ binds each formal parameter $p_i \in \text{Params}(f)$ to the sanitization status of its corresponding actual argument:

$$
\mathcal{S}^{\mathcal{V}}(p_i) = \mathcal{S}^{\mathcal{V}}(\text{ARG}^i_v) \quad \text{for } i = 1, \ldots, n
$$

This context is propagated throughout the analysis of function $f$'s body, ensuring that parameter sanitization status reflects the calling context.

\paragraph{Fixpoint Computation and Soundness}
Sanitization is a \emph{must} property, so we compute the greatest fixpoint of
$\mathcal{S}^{\mathcal{V}}$: results are memoized by $(v, \pi^{\mathcal{V}}, \Phi)$,
and a node revisited while its own evaluation is in progress (a loop-carried
dependency or recursion) is provisionally assigned $1$, which ensures termination.
Like most static analyzers for dynamic languages, \sysname{} is \emph{soundy}~\cite{LivshitsSoundiness15}: it handles the core language soundly and makes explicit choices for features that are hard to model statically. Table~\ref{tab:php-features} lists the PHP features known to affect static analysis~\cite{HillsPHPFeatures13, DahseRIPS14} and how we treat each.

\begin{figure}[h!]
    \centering
    \includegraphics[width=.9\linewidth]{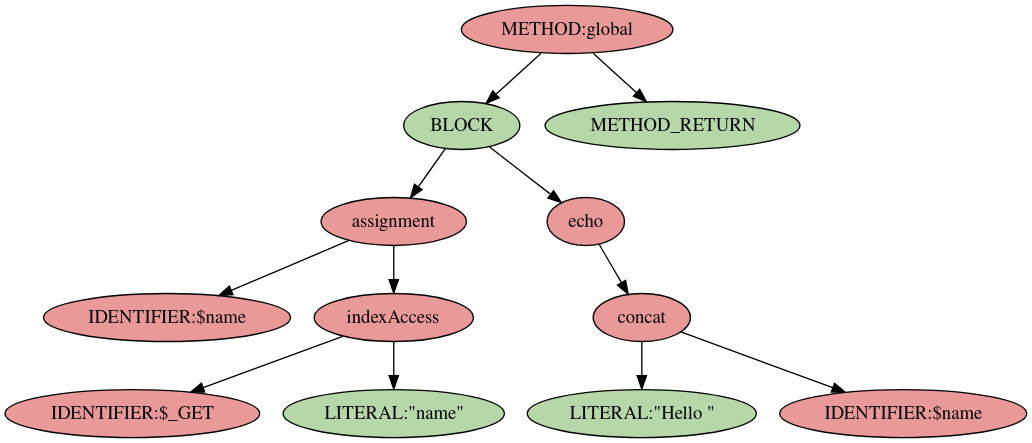}
    \caption{Unsanitized code's augmented AST}
    \label{fig:unsanAST}
\end{figure}

\paragraph{Example Walkthrough}
As illustration, we apply the sanitization algorithm above to a simple PHP code that retrieves the $name$ parameter from an HTTP request and displays it back: a simple XSS vulnerability.
\begin{lstlisting}[language=PHP]
<?php
    $name = $_GET['name'];
    echo("Hello ". $name);
?>
\end{lstlisting}
Figure~\ref{fig:unsanAST} shows the augmented AST produced: sanitized nodes are colored in green and unsanitized ones red. As expected, the sink function ($echo$) is unsanitized.

\begin{figure}[h!]
    \centering
    \includegraphics[width=.9\linewidth]{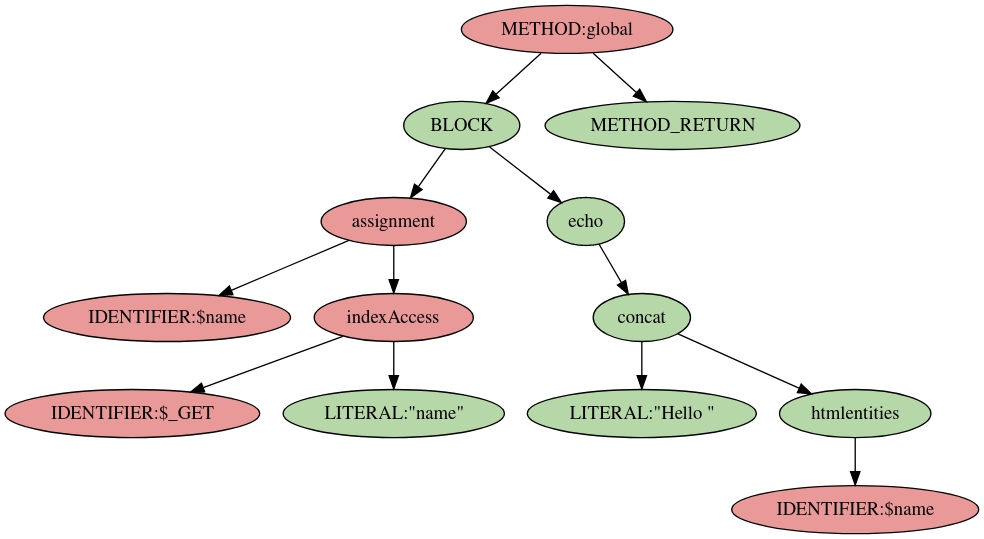}
    \caption{Sanitized code's augmented AST}
    \label{fig:sanAST}
\end{figure}

In contrast, the following code sanitizes user input before displaying it. Figure~\ref{fig:sanAST} shows how the $htmlentities$ function sanitizes the sink function, preventing the XSS vulnerability.
\begin{lstlisting}[language=PHP]
<?php
    $name = $_GET['name'];
    echo("Hello ". htmlentities($name));
?>
\end{lstlisting}
This sanitization layer enables \sysname{} to tag nodes as either clean or tainted. It directly addresses \textbf{Challenge 1}, ensuring that sanitization is correctly interpreted and that this context is preserved throughout the data flow of the code. These tags are propagated and used to add further constraints to the data flows that reach database computations, where deeper insights into both direct and indirect data flows are required. This includes the use of query and table-level constraints to improve the precision of vulnerability detection.

\textbf{Database Integration}.
The main challenge with establishing a sound relationship between code and data is that scripts sharing persistent storage can appear as functionally separate within the CPG. For example, in the PHP code of Listing \ref{lst:input} and \ref{lst:display}, data from \textit{input.php} can easily reach \textit{display.php}: independent code snippets can share data through the database without explicit data dependency edges.

To address this limitation, \sysname{} treats persistence as a first-class part of taint propagation. It parses the database schema and the SQL statements in code to (i) classify \emph{query safety} under schema- and sanitization-aware constraints and (ii) introduce \emph{database dependency edges} that reconnect write operations (INSERT or UPDATE) to read operations (SELECT) when both sides carry unsafe data. This augmentation targets \textbf{Challenge 2} by enabling cross-script vulnerability discovery, infeasible without explicitly modeling the database.

Since SQL statements in applications interleave constants, variables, and function calls, \sysname{} performs a lightweight query parsing step to extract: (1) the target table, (2) the referenced columns, and (3) the CPG value nodes that supply user-controlled inputs. Concretely, the parser tokenizes query strings, normalizes common syntactic variations, and applies pattern- and clause-aware extraction rules to recover $(t_q,\mathcal{C}_q,\mathcal{V}_q)$. The extracted metadata is then attached to the corresponding SQL call nodes in the CPG and used to drive the safety and dependency rules below.

\begin{definition}[Query Parsing]
For an SQL query $q$, we define
\[
\text{PARSE}(q) = (t_q, \mathcal{C}_q, \mathcal{V}_q)
\]
where
\[
\begin{array}{rcl}
    t_q & : & \text{target table ($t_q \in T$)}\\[4pt]
    \mathcal{C}_q & : & \text{affected columns ($\mathcal{C}_q \subseteq C$)}\\[4pt]
    \mathcal{V}_q & : & \text{CPG value nodes ($\mathcal{V}_q \subseteq V$)}
\end{array}
\]
\end{definition}

Note that the above definition implies returning all columns $\mathcal{C}$ of a table $t$ if the wildcard symbol (*) is used.

\paragraph{Query Safety Analysis} We analyze the safety of SQL statements extracted from the application code as follows.

Following the zero trust principle, \textbf{SELECT queries} $q$ are treated as unsanitized unless a safety guarantee can be inferred: either the selected columns $\mathcal{C}_q$ are type safe or the query applies a safe SQL function.

$$\mathcal{Q}^{\text{SELECT}}(q) = \begin{cases}
\text{SAFE} & \substack{\text{if } \forall c \in \mathcal{C}_q : \tau(t_q, c) = 1 \vee c \in \Phi_{\text{SQL}}} \\
\text{UNSAFE} & \text{otherwise}
\end{cases}$$

where $\Phi_{\text{SQL}} = \{\text{COUNT}, \text{SUM}, \text{LENGTH}, \ldots\}$ are type-safe built-in SQL functions.

\textbf{INSERT or UPDATE queries} $q$ are safe if the values inserted into unsafe columns are properly sanitized, ensuring that tainted data doesn't leak into persistent storage.

$$\mathcal{Q}^{\text{INSERT}}(q) = \begin{cases}
\text{SAFE} & \substack{\text{if } \{c \in \mathcal{C}_q : \tau(t_q, c) = 0\} = \emptyset} \\
\bigwedge_{v \in \mathcal{V}_{\text{unsafe}}} \mathcal{S}^{\mathcal{V}}(v) & \text{otherwise}
\end{cases}$$

where $\mathcal{V}_{\text{unsafe}} = \{v_i \in \mathcal{V}_q : c_i \in \mathcal{C}_q \wedge \tau(t_q, c_i) = 0\}$ and $\mathcal{S}^{\mathcal{V}}(v)$ is the sanitization status of the node.
If writings to storage are sanitized, then no proper unsanitized path can be found from a malicious input to the write query. Alternatively, if downstream sanitization or escaping of the values is applied to read queries, no unsanitized path can reach the sensitive sink. Either way, no vulnerable path would be reported.

\begin{definition}[Database Dependency Edge]
We add a database dependency edge from an \texttt{INSERT} or \texttt{UPDATE} query $q_1$ to a \texttt{SELECT} query $q_2$ if:
\[
\begin{array}{rcl}
    1.&\text{PARSE}(q_1) = (t_1, \mathcal{C}_1, \mathcal{V}_1) \ \text{and} \ \text{PARSE}(q_2) = (t_2, \mathcal{C}_2, \mathcal{V}_2) \\[6pt]
    2. & t_1 = t_2 \quad \text{(same table)} \\[6pt]
    3. & \mathcal{C}_1 \cap \mathcal{C}_2 \neq \emptyset \quad \text{(overlapping columns)}\footnotemark \\[6pt]
    4. & \mathcal{Q}^{\text{INSERT}}(q_1) = \text{UNSAFE} \\[4pt]
       & \quad\;\; \text{and } \mathcal{Q}^{\text{SELECT}}(q_2) = \text{UNSAFE}
\end{array}
\]
\footnotetext{A limitation of this definition is that it misses the edge case where only safe columns overlap}
\end{definition}

This creates additional data dependency edges $E_{\text{DB}}$ in the augmented CPG: $G' = (V, E \cup E_{\text{DB}}, \lambda, \mu)$, enabling taint flow detection across database boundaries even when no direct code dependencies exist between separate scripts. The benefit of this type of graph augmentation is two-fold:
\begin{enumerate}
    \item \textbf{Discovery of further vulnerable paths}:
The database dependency edges allow us to traverse sink functions across the database beyond the traditional CPG.
    \item \textbf{Reduction of false positives}:
Incorporating the constraints imposed by the database schema will rule out safe queries, decreasing the number of false positives.
\end{enumerate}

Database-aware taint analysis is applied to all taint-style vulnerabilities except second-order SQL injections. While its most popular form is persistent XSS, it is just as feasible for other vulnerability classes. For example, an input source can be saved to the database before being executed by the operating system: a stored form of command injection.

\paragraph{Example Walkthrough}
The SELECT statement in Listing \ref{lst:display} will be labeled as unsafe because one of its relevant columns (comments) is of an unsafe type as specified by its database schema. In Listing \ref{lst:input}, the INSERT statement will also be labeled as unsafe since the value inserted into the comments column isn't sanitized before it was added to the query command ({\tt \$comment}). 

This concludes the CPG augmentation part in \sysname{} as illustrated in Figure \ref{fig:overview}. Compared to the SOTA, these new enhancements enrich the CPG, ensuring backward traversal can uncover vulnerabilities that span both code and persistent state.

\textbf{Object-aware Data Flow Analysis}.
A key limitation of many CPG-based taint analyses on object-oriented code is the lack of explicit, field-level dependencies across method boundaries. In practice, security-relevant values are frequently stored in object fields, initialized in constructors, and modified in helper methods (including static utilities and dynamically dispatched calls). When these field updates are not represented as dependencies, taint propagation either breaks (false negatives) or becomes overly conservative (false positives). To address this gap (Challenge~3), \sysname{} augments the CPG with object dependency relations derived from a context-sensitive reaching-definitions analysis that tracks field writes and resolves object aliasing across call boundaries.

\begin{definition}[Object-aware Reaching Definition Domain]
The analysis operates over the domain of definition sets with calling context:
\[
\mathcal{D} = \mathcal{P}(V \times \text{Context})
\]
where
\[
\begin{array}{rcl}
    V & : & \text{CPG node}\\[6pt]
    Context & : & \text{CallStack} \times \text{VarStack} \\
    CallStack & : & 
        \begin{minipage}[t]{0.7\linewidth}
        call stack $[c_1, c_2, \ldots, c_n]$ with $c_i \in V$ caller nodes in reverse chronological order
        \end{minipage}
    \\[12pt]
    VarStack & : & 
        \begin{minipage}[t]{0.7\linewidth}
        variable stack $[v_1, v_2, \ldots, v_n]$ with $v_i$ representing variable names of the same object across different scopes
        \end{minipage}
\end{array}
\]
\end{definition}

The call and variable stacks are used to store procedure calls and variable names. When the analysis reaches a function call, the caller's node is added to the call stack before switching to the callee's body. The variable name stack follows the same logic for handling different parameter names referring to the same object or switching to the reserved keyword \texttt{this} for constructors and member functions.

For node $v$ and variable name $x$, we compute the context-sensitive reaching definition function:

\[
\mathcal{R}^{\mathcal{O}}(v, x, \kappa) = \mathcal{T}^{\mathcal{O}}\!\left(v, x, \{\mathcal{R}^{\mathcal{O}}(u, x, \kappa') : u \in \text{PRODUCERS}(v)\}, \kappa \right)
\]

where $\mathcal{T}^{\mathcal{O}}$ is the object-aware transfer function and $\kappa'$ represents the appropriate context for each predecessor.

The transfer function $\mathcal{T}^{\mathcal{O}}: V \times \text{String} \times \mathcal{P}(\mathcal{P}(V \times \text{Context})) \times \text{Context} \to \mathcal{P}(V \times \text{Context})$ processes nodes based on their types:

\textbf{Assignment nodes} create new definitions:
\begin{align}
&\text{MATCH}_{\text{assignment}}(v, x) \Rightarrow \mathcal{T}^{\mathcal{O}}(v, x, \_, \kappa) = \{(v, \kappa)\}
\end{align}

\textbf{Field access operations} require object-aware analysis across method boundaries:
\begin{align}
&\text{MATCH}_{\text{fieldAccess}}(v, x) \Rightarrow \nonumber \\
&\mathcal{T}^{\mathcal{O}}(v, x, D, \kappa) = \text{ComputeFieldDefs}(v, x, \kappa)
\end{align}
Here, $\text{ComputeFieldDefs}$ resolves definitions across object instances and method boundaries.

\textbf{Constructor and method calls} require context-sensitive inter-procedural analysis:
\begin{align}
&\text{MATCH}_{\text{call}}(v, x) \Rightarrow \nonumber \\
&\mathcal{T}^{\mathcal{O}}(v, x, D, \kappa) = \mathcal{R}^{\mathcal{O}}(\text{ReturnBlock}(\text{Callee}(v)), x', \kappa')
\end{align}
Here $x'$ is the parameter name in the callee and $\kappa'$ is the updated calling context.

\textbf{Method parameters} connect to their corresponding arguments at call sites:
\begin{align}
&\text{MATCH}_{\text{parameter}}(v, x) \Rightarrow \nonumber \\
&\mathcal{T}^{\mathcal{O}}(v, x, D, \kappa) = \bigcup_{c \in \text{CallSites}(v)} \mathcal{R}^{\mathcal{O}}(\text{Argument}_i(c), x, \kappa_c)
\end{align}
where $\text{Argument}_i(c)$ is the $i^{th}$ argument of call $c$ and $\kappa_c$ its calling context.

\textbf{Control flow boundaries} handle interprocedural returns:
\begin{align}
\text{PRODUCERS}(v) &= \emptyset \\ \Rightarrow \nonumber 
\mathcal{T}^{\mathcal{O}}(v, x, D, \kappa) &= \begin{cases}
\emptyset \quad \text{if} \quad \text{CallStack}(\kappa) = \emptyset \\
\mathcal{R}^{\mathcal{O}}(\text{CallSite}(\kappa), x, \text{PopContext}(\kappa))
\end{cases}
\end{align}

\textbf{Regular data flow} propagates existing definitions:
\begin{align}
&\text{otherwise} \Rightarrow \mathcal{T}^{\mathcal{O}}(v, x, D, \kappa) = \bigcup_{(d, \kappa_d) \in D} \{(d, \kappa_d)\}
\end{align}

Sanitization and database augmentations ensure that our traversal accounts for inter-procedural object behaviors.

\textbf{Backward Traversal}.
With sanitization, database, and object links established, we now address the need for robust data flow analysis: tracing tainted sinks back to their sources. Prior works such as~\cite{YamaguchiCPG14, BackesPHP17} introduced backward traversal to detect vulnerable paths. Our main contribution in \sysname{} is to strengthen this mechanism through the integration of data dependency walks in object-oriented contexts.

\sysname{}'s backward traversal (Algorithm~\ref{alg:backwardTraversal}) is an iterative breadth-first search: it expands the end node (a sink of vulnerability $v$) through data dependency edges until it reaches start nodes (attacker-controlled sources). Each round extends every path by the reaching definitions of its last node; for a method parameter, these are the corresponding arguments at its call sites, and field accesses are resolved in the object context $ctx(path)$ of the path. Definitions already visited are skipped to prevent cycles, and definitions sanitized for $v$ end the path. The first path reaching each start node is recorded, and the traversal terminates once no path can be extended or a node budget $B$ is exceeded. This traversal mechanism forms the core engine for identifying candidate vulnerable paths.

\begin{algorithm}[t!]
\scriptsize
\caption{Backward Traversal}
\label{alg:backwardTraversal}
\begin{algorithmic}[1]
\Procedure{BackwardTraversal}{$endNode$, $startNodes$, $v$}
  \State $paths \gets [[endNode]]$; \quad $found \gets \{\}$
  \Loop
    \State $visited \gets \bigcup_{p \in paths} p$
    \State $output \gets []$; \quad $extended \gets false$
    \ForAll{$path$ in $paths$}
      \State $n \gets path.last$
      \If{$n$ is $MethodParameter$}
        \State $defs \gets \{c.arg[n.index] \mid c \in n.method.callSites\}$
      \Else
        \State $defs \gets \Call{ReachingDefs}{n, ctx(path)}$
      \EndIf
      \State $defs \gets \{d \in defs \mid d \notin visited \wedge \neg sanitized_{v}(d)\}$
      \If{$defs.empty$}
        \State $output.\text{append}(path)$
      \Else
        \State $extended \gets true$
        \ForAll{$d$ in $defs$}
          \State $output.\text{append}(path + d)$
        \EndFor
      \EndIf
    \EndFor
    \ForAll{$s \in startNodes \setminus found$}
      \If{$\exists p \in output : s \in p$}
        \State $found[s] \gets$ first such $p$
      \EndIf
    \EndFor
    \If{$\neg extended$ \textbf{or} $|visited| > B$}
      \State \Return $found$
    \EndIf
    \State $paths \gets output$
  \EndLoop
\EndProcedure
\end{algorithmic}
\end{algorithm}

\textbf{Vulnerable Sink Discovery and CVE Matching}.
To systematically identify vulnerable paths, Algorithm~\ref{alg:sinkDisc} detects vulnerable sinks as sensitive functions that can be reached by unsanitized, attacker-controlled data. Source and sink nodes are language-specific configurations as defined in Definitions \ref{def:source-def} and \ref{def:vuln-config} respectively.

\algrenewcommand{\algorithmiccomment}[1]{\hfill{\color{gray}$\triangleright$~#1}}
\newcommand{\San}{\mathcal{S}^{\mathcal{V}}}

\begin{algorithm}[t!]
\scriptsize
\caption{Vulnerable Sink Discovery}
\label{alg:sinkDisc}
\begin{algorithmic}[1]
\Procedure{GetVulnerablePaths}{$G$, $\mathcal{V}\!=\!(\Phi,\Sigma)$, $src$}
  \State $sinks \gets \texttt{FIND\_NODES}(G, \Sigma)$
  \State $A \gets \{a \in \mathit{args}(sinks) \mid \San(a) = 0\}$ \Comment{unsanitized}
  \State $P \gets \emptyset$
  \ForAll{$a \in A$, $o \in src$}
    \Statex \hspace{\algorithmicindent}\hspace{\algorithmicindent}{\color{gray}$\triangleright$ BFS; skips nodes with $\San = 1$}
    \State $p \gets \texttt{BackwardTraversal}(G, a, o, \San)$
    \If{$p \neq \bot \wedge \neg\texttt{HAS\_DB\_CALL}(p)$}
      \State $P \gets P \cup \{p\}$
    \EndIf
  \EndFor
  \If{$\mathcal{V} \neq \text{SQLi}$} \Comment{stored flows}
    \State $P \gets P \cup \texttt{InterQueryPaths}(G, A, src, \San)$
  \EndIf
  \State \Return $\texttt{DEDUP\_BY\_ENDPOINTS}(P)$ \Comment{(source, arg)}
\EndProcedure
\end{algorithmic}
\end{algorithm}

As a final step, \sysname{} iterates over reported vulnerable paths and cross-references them with public CVEs. Matching leverages NLP techniques including tokenization, stopword filtering, and regex-based pattern extraction to identify file names, vulnerable functions, application versions, and code parameters from unstructured CVE descriptions. We employ multi-layered semantic analysis with vulnerability type classification, version comparison, and parameter-based code matching using substring identification with word boundary detection. This best-effort approach allows us to filter reported vulnerable paths into those associated with a publicly available exploit to confirm exploitability, addressing the final challenge and ensuring the precision and relevance of the results.

\section{Evaluation}\label{sec: eval}

We evaluate \sysname{} across
synthetic unit tests, benchmark datasets, and real-world PHP applications.
The following questions guide our evaluation:
\begin{enumerate}
    \item \textbf{RQ1 (Effectiveness):} 
    How accurately does \sysname{} detect taint-style vulnerabilities on labeled benchmark datasets compared to state-of-the-art static analysis tools and learning-based approaches?

    \item \textbf{RQ2 (Real-World Applicability):} 
    How effective is \sysname{} at detecting known and previously unknown vulnerabilities in real-world PHP applications?

    \item \textbf{RQ3 (Precision and Practicality):} 
    Does \sysname{} reduce false positives and redundant reports compared to existing tools when rediscovering known vulnerabilities?

    \item \textbf{RQ4 (Component Contribution):} 
    How do individual components of \sysname{}’s analysis pipeline contribute to detection coverage and precision?
\end{enumerate}

To answer \textbf{RQ1}, we evaluate \sysname{} on handcrafted unit tests and the
SARD benchmark~\cite{nist2023sard}, comparing it against established static analyzers (Pixy~\cite{1624016}, Kave~\cite{KAVE}, and PHPCorrector~\cite{PHPCorrec}) using standard classification metrics.
To address \textbf{RQ2} and \textbf{RQ3}, we evaluate \sysname{} on large-scale,
real-world PHP applications in with respect to publicly reported CVEs and the state-of-the-art.
Finally, to answer \textbf{RQ4}, we conduct a detailed ablation study isolating
the impact of each major component in \sysname{}’s pipeline.

\subsection{Dataset and Experimental Setup}
\label{sec:steup}
\textbf{Dataset}. Our evaluation covers diverse codebase complexities and application contexts. We sample 10,000 PHP snippets from the SARD dataset~\cite{nist2023sard} and select 19 popular GitHub-hosted PHP applications ranging from simple web tools to mature systems. This benchmark enables a rigorous, fair comparison against state-of-the-art tools, including FIXX~\cite{thimmaiahfixx}, Zipper \cite{zipper},Tchecker\cite{10.1145/354860} and Yama \cite{Yama}. Targets and baselines were selected based on three criteria: (i) public availability and reproducibility, (ii) representative coverage of CPG/taint static analyzers evaluated in prior work, and (iii) the presence of ground-truth labels (SARD) or verifiable CVEs.

\textbf{Environment}. The evaluation environment consists of an Ubuntu 20.04 LTS machine with 10 cores (3.1 GHz each) and 32 GB of RAM. Joern is used to parse the application's code into a CPG. \sysname{} augments the CPG with relevant sanitization and database query labeling, and perform vulnerable sink discovery through backward traversal. Python scripts are used to cross-reference the vulnerable paths produced with the CVE database. We also manually verify a small sample of the paths to group them into zero-days or false positives.

\textbf{Evaluation Stages}. \sysname{} is evaluated at 3 different levels:
\begin{enumerate}
    \item \textbf{Handwritten Unit Tests}: We design targeted unit tests capturing complex coding logic to guide the development of \sysname{}'s modules and algorithms and capture complex taint flow in isolated, controlled environments. These tests include both positive (vulnerable) and negative (secure) cases, allowing us to detect any regression after configuration or algorithmic updates and assess detection accuracy and false positive rates.
    \item \textbf{Evaluation Dataset}: We leverage SARD, a large-scale benchmark containing real-world and synthetic vulnerability samples mapped to Common Weakness Enumeration (CWE) categories (Table \ref{tab:targeted-CWE} in the Appendix). We sample from the dataset to improve the language-specific configuration for better coverage and evaluate it on a separate holdout sample. This enables systematic benchmarking of \sysname{} against controlled vulnerabilities across simple code snippets for PHP.
    \item \textbf{Full Application Code}:  \sysname{} is used on complete open-source and enterprise applications to assess its performance on real-world codebases. No application specific finetuning happens at this stage. We measure scalability and accuracy in identifying vulnerabilities , ensuring that \sysname{} can be leveraged by practitioners and researchers.
\end{enumerate}
\vspace{-0.3cm}

\subsection{RQ1: Effectiveness on Benchmarks}

We first evaluate \sysname{} on handcrafted PHP unit tests designed to exercise complex taint flows, including validation guards, pass-by-reference propagation, inter-procedural flows, second-order database flows, object-aware dependencies and file inclusion. \sysname{} correctly classified all test cases as vulnerable or secure.

We further evaluate \sysname{} on the SARD benchmark for PHP, comparing it
with KAVe~\cite{KAVE}, WAP~\cite{WAP}, Pixy~\cite{1624016}, and
PHPCorrector~\cite{PHPCorrec}. As shown in Table~\ref{tab:php_sard_full},
\sysname{} achieves the highest overall accuracy (82.56\%), and F1 score (76.46\%), outperforming other benchmarks while maintaining a low false positive rate (18.87\%).

{\small
\begin{summarybox}
\textbf{Takeaway 1:} \sysname{} substantially outperforms existing static analyzers on labeled benchmarks, demonstrating that semantic-aware taint propagation and sanitization modeling significantly improve detection
accuracy without inflating false positives.
\end{summarybox}
}

\begin{table*}[t]
\centering
\footnotesize
\setlength{\tabcolsep}{4pt}
\newcommand{\missed}{\rlap{\,\ding{55}}}
\scalebox{0.9}{
\begin{tabular}{@{}l r l l ccccc l@{}}
\toprule
 & & & & \multicolumn{5}{c}{\textbf{Reported paths} (\ding{55}\,=\,CVE not rediscovered)} & \\
\cmidrule(lr){5-9}
\textbf{Application} & \textbf{LOC} & \textbf{CVE} & \textbf{Type} &
\textbf{FIXX} & \textbf{ZIPPER} & \textbf{Yama} & \textbf{TChecker} & \textbf{\sysname{}} &
\textbf{Zero-days (CVE-2025-)} \\
\midrule

SeoPanel                      & 255.0k & CVE-2021-3002  & XSS  &  2\missed &        28 &  0\missed &        13 &        13 & -- \\
\addlinespace[2pt]

Collabtive                    & 180.4k & CVE-2021-3298  & XSS  &  9\missed &         6 &  0\missed &         1 &  0\missed & -- \\
                              &        & CVE-2024-46240 & XSS  &         9 &         6 &  0\missed &         1 &  0\missed &  \\
                              &        & CVE-2024-48706 & XSS  &         9 &         6 &  0\missed &         1 &  0\missed &  \\
                              &        & CVE-2024-48707 & XSS  &         9 &         5 &  0\missed &         1 &  0\missed &  \\
                              &        & CVE-2024-48708 & XSS  &         9 &         5 &  0\missed &         0 &  0\missed &  \\
\addlinespace[2pt]

osCommerce CE Phoenix         &  59.8k & CVE-2020-12058 & XSS  &  0\missed &       345 &  0\missed &        32 &        22 & -- \\
\addlinespace[2pt]

Clansphere CMS                &  54.3k & CVE-2021-27310 & XSS  & 20\missed &        17 &  0\missed &       171 &         3 & -- \\
\addlinespace[2pt]

Fantastic Blog CMS            &  24.7k & CVE-2022-28512 & SQLi &         9 &       120 &  0\missed &        26 &         1 & 65337 \\
                              &        & CVE-2021-26231 & SQLi &        42 &       120 &  0\missed &        25 &         1 &  \\
                              &        & CVE-2021-26224 & XSS  &  0\missed &       107 &  0\missed &  0\missed &         1 &  \\
\addlinespace[2pt]

Engineers Online Portal       &  15.6k & CVE-2023-5283  & SQLi &       258 &       171 &  0\missed &       197 &        69 & -- \\
                              &        & CVE-2023-5276  & SQLi &        52 &       170 &  0\missed &       196 &        73 &  \\
\addlinespace[2pt]

Hospital Mgmt.\ System        &   9.4k & CVE-2021-39411 & XSS  &         6 &        20 &  0\missed &        30 &         6 & 65340, 69942--69945, \\
                              &        & CVE-2024-46237 & XSS  &        10 &        20 &  0\missed &        30 &        11 & 69949 \\
                              &        & CVE-2024-46238 & XSS  &         9 &        19 &  0\missed &        30 &         5 &  \\
                              &        & CVE-2024-46239 & XSS  &         9 &        19 &  0\missed &        30 &        15 &  \\
\addlinespace[2pt]

Advocate Office Mgmt.         &     9k & CVE-2024-9328  & SQLi &        21 &        58 &  0\missed &        40 &        24 & 67406 \\
\addlinespace[2pt]

CASAP Autom.\ Enrollment      &   7.7k & CVE-2021-3294  & XSS  &        21 &  0\missed &  0\missed &  0\missed &        15 &
67403--67405, 67407, 67408, \\
                              &        & CVE-2021-26226 & SQLi &  0\missed &       212 &        49 &       135 &        17 & 65343$^\dagger$, 69939$^\dagger$ \\
\addlinespace[2pt]

Tailor Mgmt.\ System          &   7.7k & CVE-2021-40260 & XSS  &        86 &  0\missed &  0\missed &  0\missed &        11 & 69941, 69947 \\
                              &        & CVE-2020-36073 & SQLi &       130 &        30 &  0\missed &        63 &         7 &  \\
\addlinespace[2pt]

Fruits Bazar                  &   6.4k & CVE-2022-34989 & SQLi &         4 &        24 &  0\missed &        42 &         8 & 65336, 65341 \\
                              &        & CVE-2022-30478 & SQLi &         1 &        24 &  0\missed &        42 &        14 &  \\
\addlinespace[2pt]

Blood Bank System             &   4.8k & CVE-2024-9327  & SQLi &         5 &        48 &        10 &        19 &        23 & 65342 \\
                              &        & CVE-2024-9804  & SQLi &        84 &        47 &         9 &        19 &        22 &  \\
\addlinespace[2pt]

Covid19 TMS                   &   4.1k & CVE-2024-53604 & SQLi &        42 &        21 &  0\missed &        20 &         5 & -- \\
                              &        & CVE-2024-53603 & SQLi &        43 &        20 &  0\missed &        20 &        11 &  \\
\addlinespace[2pt]

Dairy Farm Mgmt.              &     3k & CVE-2023-41593 & XSS  &        33 &        22 &  0\missed &        17 &        12 & -- \\
                              &        & CVE-2024-46241 & XSS  &         9 &        22 &  0\missed &        16 &        14 &  \\
\addlinespace[2pt]

CodeAstro Membership          &   2.6k & CVE-2024-25868 & XSS  &        38 &        19 &  0\missed &  0\missed &         2 &
69930, 69931, \\
                              &        & CVE-2024-46236 & XSS  &         5 &        19 &  0\missed &  0\missed &         3 & 69933--69938 \\
                              &        & CVE-2024-48709 & XSS  &         9 &        18 &  0\missed &  0\missed &         2 &  \\
                              &        & CVE-2024-46472 & SQLi &        44 &        26 &  0\missed &        25 &         3 &  \\
                              &        & CVE-2024-2333  & SQLi &        17 &        26 &  0\missed &        25 &         1 &  \\
\addlinespace[2pt]

Modern Loan Mgmt.             &   2.2k & CVE-2024-0900  & SQLi &        40 &        23 &        15 &        59 &        15 & 69946, 69948 \\
\addlinespace[2pt]

Daily Expense Tracker         &   1.7k & CVE-2020-10106 & SQLi &        10 &        20 &         3 &        30 &         4 & -- \\
                              &        & CVE-2021-26304 & XSS  &        27 &        32 &  0\missed &  0\missed &         6 &  \\
\addlinespace[2pt]

Visitor Management            &     1k & CVE-2024-22983 & SQLi &        21 &  0\missed &  0\missed &         6 &         2 & -- \\
                              &        & CVE-2020-25760 & SQLi &         0 &  0\missed &  0\missed &         6 &  0\missed &  \\
                              &        & CVE-2020-25761 & SQLi &        21 &  0\missed &  0\missed &         5 &         4 &  \\
\addlinespace[2pt]

Keerti                        &    968 & CVE-2024-1700  & XSS  &        11 &  0\missed &  0\missed &  5\missed &  5\missed & -- \\

\midrule
\multicolumn{4}{@{}l}{\textbf{Total}} &
1{,}184 & 1{,}895 & 86 & 1{,}378 & 450 & \textbf{30} \\

\multicolumn{4}{@{}l}{\textbf{CVEs rediscovered} (of 42)} &
36 & 36 & 5 & 34 & 35 & \\

\bottomrule
\end{tabular}}

\caption{Comparison with FIXX~\cite{thimmaiahfixx}, ZIPPER~\cite{zipper},
Yama~\cite{Yama}, and TChecker~\cite{10.1145/354860} on 19 PHP web
applications with 42 known CVEs. Each tool cell gives the number of paths
the tool reports for that CVE; \ding{55} marks not rediscovered CVE. The \emph{Zero-days} column lists the CVE-2025 identifiers discovered by \sysname{} and accepted as CVEs.
$^\dagger$Reserved}

\label{tab:evaluation-results-sota}
\end{table*}
\begin{table}[t!]
    \centering
    \renewcommand{\arraystretch}{1.2}
    \scalebox{0.9}{
    \begin{tabular}{|c|c|c|c|c|c|c|}
        \hline
        \textbf{Language} & \textbf{Approach} & \textbf{Acc} & \textbf{F1} & \textbf{Prec} & \textbf{Rec} & \textbf{FPR} \\
        \hline
        \multirow{5}{*}{PHP\textsuperscript{1}} 

        & KAVe \cite{KAVE} & 56.00 & 38.00 & 60.00 & 27.00 & 17.00 \\
        & WAP \cite{WAP}  & 43.00 & 33.00 & 37.00 & 29.00 & 44.00 \\
        & Pixy \cite{1624016} & 66.00 & 68.00 & 64.00 & 73.00 & 41.00 \\
        & PHPCorrector \cite{PHPCorrec} & 53.00 & 3.00 & \textbf{86.00} & 2.00 & \textbf{0} \\
        & \sysname{} & \textbf{82.56} & \textbf{76.46} & 73.84 & \textbf{85.15} & 18.87 \\

        \hline
    \end{tabular}}
    \caption{Performance comparison of \sysname{} with previous vulnerability analysis tools on the SARD benchmark.}
    \label{tab:php_sard_full}
    \vspace{-0.5cm}
\end{table}

\subsection{RQ2: Effectiveness on CVE and Exploit Detection}

We evaluate \sysname{} on 19 real-world PHP applications of varying size and
complexity. For each application, we compare detected vulnerable paths against
publicly disclosed CVEs and manually validate newly discovered issues.
Table~\ref{tab:evaluation-results-sota} summarizes the results.

Across all applications, \sysname{} successfully re-discovered all but seven known
CVEs and uncovered 30 confirmed zero-day vulnerabilities: 27 SQL injections and 3 XSS across 9 distinct applications.

{\small
\begin{summarybox}
\textbf{Takeaway 2:} \sysname{} demonstrates strong real-world effectiveness by reliably
rediscovering known CVEs and uncovering previously unknown vulnerabilities across
diverse PHP applications, confirming practical utility beyond benchmark
datasets.
\end{summarybox}
}

\subsection{RQ3: Precision and Vulnerability Report Quality}
We compare \sysname{} against FIXX~\cite{thimmaiahfixx},
ZIPPER~\cite{zipper}, Yama~\cite{Yama}, and
TChecker~\cite{10.1145/354860} on the same 19 PHP applications and
42 known CVEs (Table~\ref{tab:evaluation-results-sota}). \sysname{}
rediscovers 35 CVEs with 450 CVE-matched paths, compared with
36/1,184 for FIXX, 36/1,895 for ZIPPER, 5/86 for Yama and 34/1,378 for TChecker.
Overall, \sysname{} maintains comparable CVE rediscovery while
substantially reducing CVE-matched paths relative to FIXX, ZIPPER,
and TChecker, and covering substantially more CVEs than Yama.

Table~\ref{tab:evaluation-results-sota} reports paths associated with
known CVEs, and Table~\ref{tab:database_approach} reports all
candidate paths produced by \sysname{} across the seven supported
vulnerability classes, including false positives.
{\small
\begin{summarybox}
\textbf{Takeaway 3:} By pruning infeasible and redundant taint paths, \sysname{}
reduces analyst triage effort while preserving vulnerability recall, improving the
practical usability of static vulnerability detection.
\end{summarybox}
}

\begin{table}[t]
\centering
\scriptsize
\setlength{\tabcolsep}{2.2pt}
\resizebox{\columnwidth}{!}{
\begin{tabular}{lcccccccc}
\toprule
\shortstack{\textbf{Web}\\\textbf{Application}} &
\shortstack{\textbf{Execution}\\\textbf{Time (s)}} &
\shortstack{\textbf{Code}\\\textbf{Injection}} &
\shortstack{\textbf{Command}\\\textbf{Execution}} &
\shortstack{\textbf{File}\\\textbf{Inclusion}} &
\shortstack{\textbf{Session}\\\textbf{Fixation}} &
\shortstack{\textbf{File}\\\textbf{Access}} &
\shortstack{\textbf{SQL}\\\textbf{Injection}} &
\textbf{XSS} \\
\midrule

Advocate Office             & 2.42   & 0 & 0   & 0  & 0   & 0   & 142 & 443 \\
Automated Enrollment        & 0.74   & 0 & 0   & 0  & 0   & 0   & 537 & 4311 \\
Blood Bank System           & 0.06   & 0 & 0   & 0  & 0   & 0   & 127 & 36 \\
ClanSphere                  & 8.40   & 4 & 0   & 21 & 440 & 460 & 306 & 1287 \\
CodeAstro                   & 0.21   & 0 & 0   & 0  & 0   & 0   & 73  & 384 \\
Collabtive                  & 170.81 & 3 & 0   & 0  & 0   & 172 & 0   & 1556 \\
Covid19tms                  & 0.07   & 0 & 0   & 0  & 0   & 0   & 63  & 7 \\
Daily Expense Tracker       & 0.03   & 0 & 0   & 0  & 0   & 0   & 26  & 8 \\
Dairy Farm Management       & 0.12   & 0 & 0   & 0  & 0   & 0   & 75  & 95 \\
Fruits Bazar                & 1.59   & 0 & 20  & 0  & 0   & 0   & 263 & 298 \\
Engineers Online Portal     & 5.52   & 0 & 0   & 0  & 0   & 0   & 466 & 4373 \\
Fantastic Blog              & 34.89  & 0 & 0   & 1  & 0   & 4   & 22  & 9449 \\
Hospital Management System  & 68.74  & 0 & 0   & 0  & 0   & 18  & 151 & 695 \\
Keerti                      & 0.02   & 0 & 0   & 0  & 0   & 0   & 16  & 0 \\
Loan Management             & 0.43   & 0 & 0   & 0  & 0   & 0   & 52  & 62 \\
SEO Panel                   & 131.74 & 0 & 0   & 14 & 0   & 9   & 239 & 12488 \\
Tailor                      & 6.04   & 0 & 331 & 0  & 0   & 0   & 97  & 1021 \\
Visitor Management          & 0.13   & 0 & 0   & 0  & 0   & 0   & 41  & 61 \\
\bottomrule
\end{tabular}
}
\caption{Vulnerability detection results of \sysname{} across PHP applications for all covered vulnerability types.}
\label{tab:database_approach}
\end{table}

\begin{table}[t]
\centering
\scriptsize
\setlength{\tabcolsep}{2.0pt}
\resizebox{\columnwidth}{!}{
\begin{tabular}{llcccccccc}
\toprule
 &  &
\multicolumn{3}{c}{\shortstack{\textbf{SQL Injection}\\\textbf{Vulnerabilities}}} &
\multicolumn{4}{c}{\shortstack{\textbf{XSS}\\\textbf{Vulnerabilities}}} \\
\cmidrule(lr){3-5}
\cmidrule(lr){6-9}

\textbf{Lang.} &
\shortstack{\textbf{App}\\\textbf{Name}} &
\shortstack{\textbf{Vanilla}\\\textbf{Joern}} &
\textbf{+Dataflow} &
\shortstack{\textbf{+Saniti-}\\\textbf{zation}} &
\shortstack{\textbf{Vanilla}\\\textbf{Joern}} &
\textbf{+Dataflow} &
\shortstack{\textbf{+Saniti-}\\\textbf{zation}} &
\textbf{+Database} \\
\midrule

\multirow{7}{*}{PHP}
& Advocate Office          & 23  & 526   & 142 & 16 & 99   & 119  & 443  \\
& Blood Bank System        & 34  & 261   & 127 & 1  & 0    & 0    & 36   \\
& CodeAstro                & 24  & 234   & 73  & 21 & 11   & 11   & 384  \\
& Collabtive               & 42  & 11312 & 0   & 9  & 2403 & 1556 & 1556 \\
& Fruits Bazar             & 17  & 568   & 263 & 18 & 252  & 145  & 298  \\
& Engineers Online Portal  & 132 & 1165  & 466 & 67 & 715  & 137  & 4373 \\
& Tailor                   & 48  & 432   & 97  & 21 & 102  & 76   & 1021 \\
\bottomrule
\end{tabular}
}
\caption{Ablation study of SQL injection and XSS paths across different components of \sysname{}.}
\label{tab:ablation}
\end{table}

\subsection{RQ4: Contribution of Analysis Components}
We conduct an ablation study isolating the impact of individual components of \sysname{}, as shown in Table~\ref{tab:ablation}. Our vanilla Joern approach uses the same configuration setting as a starting point without any further enhancement. Inter-procedural and object-aware dataflow substantially increase detection coverage but introduce additional false positives. Sanitization analysis prunes up to half of the initially reported paths, reducing manual work needed for further exploit confirmation. Finally, the database aware augmentation enables detection of vulnerabilities spanning persistent storage, particularly stored XSS attacks.

Out of the 3 XSS zero-days we uncovered, CVE-2025-65342 in Blood System 1.0 is a stored XSS detected by the database-aware module as shown in Figure~\ref{fig:stored_xss}. Since the two halves share no code-level dataflow, earlier configurations see the sink but find no path to it. Table~\ref{tab:ablation}) shows a single XSS path detected by the vanilla Joern approach, a false positive unrelated to the CVE. Our schema augmentation matches both queries on table \texttt{donate} through their overlapping columns and adds the missing edge, producing 36 inter-database paths that cover all six injected fields. As for the rediscovered CVEs in Table~\ref{tab:evaluation-results-sota}, all stored XSS (e.g., CVE-2024-46236 for Code Astro, CVE-2023-41593 for Dairy Farm, etc.) are attributed exclusively to \sysname{}'s database module. Concerning the SQL injection attacks uncovered, they are attributed to our augmented dataflow module as our current implementation doesn't cover second-order SQLi attacks.

{\small
\begin{summarybox}
\textbf{Takeaway 4:} Each component of \sysname{}’s pipeline contributes meaningfully to either
detection coverage or precision, validating the design choice of combining
semantic dataflow, sanitization reasoning, and database-aware analysis.
\end{summarybox}
}

\begin{figure}[t]
\centering
\begin{tikzpicture}[
  node distance=6mm and 10mm,
  box/.style={draw, rounded corners=2pt, font=\scriptsize\ttfamily,
              minimum width=30mm, minimum height=6mm, align=center},
  file/.style={font=\scriptsize\itshape, text=gray},
  flow/.style={-{Stealth[length=2mm]}, thick},
  hop/.style={-{Stealth[length=2mm]}, thick, dashed, blue!70!black}
]
  \node[box]                     (src) {\$\_POST['city']\\don.php:9};
  \node[box, right=of src]       (ins) {INSERT INTO donate\\don.php:29};
  \node[box, below=10mm of ins]  (sel) {SELECT * FROM donate\\bbms.php:65};
  \node[box, below=10mm of src, fill=red!8] (snk) {echo "<td>\$city</td>"\\bbms.php:94};

  \draw[flow] (src) -- (ins);
  \draw[hop]  (ins) -- node[right, font=\scriptsize\itshape]{DB hop} (sel);
  \draw[flow] (sel) -- (snk);

  \node[file, above=1mm of src.north west, anchor=south west] {injection};
  \node[file, below=1mm of snk.south west, anchor=north west] {execution (any visitor)};
\end{tikzpicture}
\caption{Stored XSS in CVE-2025-65342. The dashed edge is added by our schema augmentation, which matches the \texttt{INSERT} and \texttt{SELECT} on table \texttt{donate}.}
\label{fig:stored_xss}
\end{figure}

\subsection{Limitations}
\label{app:limitations}
Being a static analysis system, \sysname{} does not execute PHP code.
Although our analysis models common PHP sources, sinks, sanitizers, object-field
flows, file inclusions, and database interactions, it remains limited by the
information that can be recovered statically. Highly dynamic PHP features, such
as runtime-generated includes, variable variables, dynamically constructed
function or method names, reflection, and code executed through \texttt{eval}-like
constructs, may be missed when their targets cannot be resolved from the source
code. Similarly, dynamically generated SQL whose table names, column names, or
query structure cannot be recovered may reduce the precision of the database-aware
analysis.

Our persistence-aware analysis models cross-script flows through the database for
stored vulnerabilities such as stored XSS, where attacker-controlled data is
written in one request and later read and emitted in another. However, we do not
claim coverage of second-order SQL injection. In particular, if tainted
data is first stored in the database and later retrieved and used to construct a
new SQL query, the current SQLi analysis doesn't connect the write and
read phases as a single exploit chain. Thus, second-order SQLi patterns that
depend on database persistence, runtime query construction, or application state
across multiple requests remain an important limitation and direction for future
work.

\section{Related Work}\label{sec:related}
We position \sysname{} with respect to previous works on taint-style vulnerability detection through CPG traversals. Static code analysis is the most popular approach to detect vulnerabilities in PHP applications~\cite{Yu2013AutomatabasedSS,Samuel,Securing,Doup,DahseRIPS14,modulevuln,KAVE,Yama,10.1145/3442381,10.1145/1250734,10.5555/2028067,10.1145/204670,10.1145/354860,1624016, 10.1145/2660267.2660363, 10.5555/2671225.2671288, DBLP:journals/jksucis/MarashdihZS23,livshits2005finding}. In particular, taint-based analysis suffers from a considerable false positive rate, requiring manual verification through detection triage. On the other hand, \sysname{} uses multiple flow analysis to infer semantically rich context to improve detection coverage while minimizing false positive detection.

\textbf{Code Property Graph Analysis}:
Backes et al.~\cite{BackesPHP17} adapted CPG traversals to PHP vulnerability detection. We extend their work with sanitization analysis, database integration, and object-dependency traversal.
NAVEX~\cite{AlhuzaliNAVEX18} combines static and dynamic analysis to generate exploits, whereas our approach is purely static.
Wi et al.~\cite{10.1145/3485447.3512235} use subgraph isomorphism over sliced CPGs to find vulnerable code. We instead analyze the full CPG to avoid the information loss that slicing introduces.
Zhao et al.~\cite{app13169240} improve PHP call graphs and taint tracking but do not model object dependencies, sanitization, or database constraints.

\textbf{Object Dependency Graph Analysis}:
Chen et al. \cite{713535} introduced Object Dependency Graph (ODG) to represent data flow across object attributes. Najumudheen et. al~\cite{10.1145/1507195.1507208} applied ODG for test coverage analysis via a call-based system dependence graph to cover dependencies, flow, call graphs, and inheritance. Li et al.~\cite{DBLP:conf/uss/LiKHC22} implemented ODG for vulnerability discovery in Node.js to resolve objects and variables across different scopes. Our approach extends previous work on ODG by integrating it with the application's CPG to enrich the context used for vulnerability detection.

\textbf{Sanitization Support}:
Weinberger et al. \cite{10.1007/978-3-642-23822-2_9} analyze inconsistencies between web framework sanitization and XSS defense, finding that most frameworks fall short. In turn, Su et al. \cite{9978990} propose a sanitizer-centric PHP code analysis, significantly reducing false positives. \sysname{} extends the sanitization support for vulnerabilities other than XSS.


\textbf{Database Integration}:  
Sadun Haq et al. \cite{haq2024lucidframeworkreducingfalse} use database constraints to reduce false positives in container scanning. We extend their approach by labeling queries at the node-level with relevant information and augment the data flow edges accordingly to achieve both improved coverage and false positive reduction.
\section{Conclusion}\label{sec: concl}

This paper presented \sysname{}, a static analysis framework that advances taint-style vulnerability discovery by enriching Code Property Graph (CPG) representations with semantic context. Through vulnerability-typed sanitization, persistence-aware dataflow tracking, and object-aware interprocedural dependency analysis, \sysname{} addresses key limitations of prior CPG-based approaches across coarse validation modeling, shallow object handling, and database blindness.

Our multi-stage evaluation shows that \sysname{} yields clear empirical benefits: outperforming state-of-the-art static analyzers on standard benchmarks, re-discovering major CVEs on PHP applications, and uncovering 30 zero-day exploits. Crucially, \sysname{} achieves high recall without significant path inflation by pruning unreachable flows and reducing redundant reports, lowering analyst triage effort. Ablation results further confirm each layer's contribution to the detection coverage and precision of the tool. Overall, \sysname{} demonstrates how essential context-rich and compositional semantic layers are for precise, scalable, and practical static taint analysis.

\section{Ethical Considerations}
In this work, we follow the {\em responsible disclosure} principle in reporting zero-days. For all newly discovered vulnerabilities, we submit the exploits to the respective vendors and to the MITRE Corporation to request CVE IDs.

When reporting to a vendor, for each zero-day, we include the description of the vulnerability with details of the vulnerability type, the part of the codebase affected (e.g., file name), proof-of-concept that demonstrates how the vulnerability is exploited, and suggestions for patching the vulnerability. 

When reporting a vulnerability to the MITRE Corporation to obtain CVE-ID, we organized the details of the vulnerabilities per application in a repository accessible only to the CVE validation team. We will continue to evaluate \sysname{} on more PHP applications and report new vulnerabilities to the vendor and MITRE as described above. 


\bibliographystyle{IEEEtran}
\bibliography{main}

\appendices
\section*{Appendix}

\subsection{Soundness of sanitization module for dynamic PHP features.}

Table~\ref{tab:php-features} shows how \sysname{} treats the dynamic PHP features in its sanitization analysis. Features resolved to $0$ can only cause false positives. Features
marked \emph{may miss} can cause false negatives.

\begin{table}[t]
\centering
\scriptsize
\caption{Treatment of dynamic PHP features in the sanitization analysis.}
\label{tab:php-features}
\begin{tabular}{p{0.30\linewidth} p{0.44\linewidth} p{0.14\linewidth}}
\hline
\textbf{Feature} & \textbf{Treatment} & \textbf{Effect} \\
\hline
Dynamic dispatch (unique target) & Resolved by name; context-sensitive analysis & Precise \\
Dynamic dispatch (multiple targets) & Unsanitized ($0$) & May over-report \\
Variable functions, callbacks (\texttt{\$f()}, \texttt{call\_user\_func}) & Default rule: conjunction of arguments & May miss \\
Magic methods (\texttt{\_\_get}, \texttt{\_\_call}, \texttt{\_\_toString}) & Not modeled; field reads unsanitized & May miss / over-report \\
Dynamic \texttt{include}/\texttt{require} & Not resolved; definitions across files not linked & May miss \\
Variable variables (\texttt{\$\$x}), \texttt{extract()}, \texttt{global} & No reaching definition; empty conjunction ($1$) & May miss \\
Reference assignment (\texttt{\$a = \&\$b}) & Only by-reference arguments are tracked & May miss \\
\texttt{eval}, \texttt{create\_function} & Modeled only as code-injection sinks & May miss \\
Untyped values (\texttt{ANY}) & Callee return type fallback; else unsafe & Conservative \\
Analysis failures (malformed CPG) & Unsanitized ($0$) & May over-report \\
Bail-out validation (\texttt{if (!valid(\$x)) die();}) & Not modeled; later reads unsanitized & May over-report \\
\texttt{do-while}, \texttt{switch} guards & No guard inferred & May over-report \\
\hline
\end{tabular}
\end{table}

\subsection{Benchmark Dataset and Reproducibility}
\label{app:benchmark-reproducibility}
We evaluate on the NIST SARD \emph{PHP test suite -- XSS, SQLi 1.0.0} (test suite \#114)~\cite{DBLP:conf/icst/SchuckertLK22}, a corpus of 248{,}592 synthetic PHP cases labelled by sanitization status in which each case combines an attacker-controlled source, a sanitization construct, a dataflow pattern and a sink. We use cases 10{,}001--20{,}000 in the order of the suite's \texttt{sarifs.json} manifest. The configuration described by \sysname{} was developed on the first 10{,}000 cases, which share no cases with the evaluation set, so no case in our results was used to tune \sysname{}. The dataset is well suited for evaluating our sanitization module: its cases include transforming sanitizers, validating predicates, type comparisons and whitelist patterns. Table~\ref{tab:targeted-CWE} lists the CWEs covered. The full list of evaluated case IDs is included in our artifact.

\begin{table}[h!]
    \centering
    \renewcommand{\arraystretch}{1.2}
    \small
    \begin{tabular}{p{1.1cm} p{2.3cm} p{3.2cm}}
        \toprule
        \textbf{CWE} & \textbf{Name} & \textbf{Description} \\
        \midrule
        CWE-89 & SQL Injection & Unsanitized input reaching a database query \\
        CWE-79 & Cross-site Scripting & Unsanitized input reaching an output sink \\
        \bottomrule
    \end{tabular}
    \caption{Targeted CWE vulnerabilities for PHP}
    \label{tab:targeted-CWE}
\end{table}
\begin{table*}[h!]
\centering
\scriptsize
\setlength{\tabcolsep}{3pt}
\renewcommand{\arraystretch}{1.05}
\caption{Benchmark source snapshots, affected versions, and local configurations.}
\label{tab:benchmark-app-versions}
\begin{tabular}{l l l l}
\toprule
\textbf{Application} & \textbf{Source Identifier} & \textbf{Affected Version} & \textbf{Local Configuration} \\
\midrule
SeoPanel & \texttt{seopanel} & 4.8.0 & \texttt{/seopanel}, MySQL \\
Collabtive & \texttt{collabtive} & 3.1 & Web root, MySQL/SQLite installer \\
osCommerce CE Phoenix & \texttt{cephoenix1050} & 1.0.5.0 & Web root, PHP 7.0--7.3, MySQL \\
ClanSphere CMS & \texttt{clansphere201144} & 2011.4 & Web root, MySQL/PostgreSQL/SQLite supported \\
FantasticBlog & \texttt{fantasticblog} & 1.0 & Web root, DB: \texttt{blog\_admin\_db} \\
Engineers Online Portal & \texttt{niamunozmonitoringsystem} & 1.0 & Web root, DB: \texttt{capstone} \\
Hospital Management System & \texttt{hospitalmanagementsystemproject4} & 4.0 & \texttt{/hospital}, DB: \texttt{hms} \\
Advocate Office Management System & \texttt{advocate} & 1.0 & \texttt{/advocate/kortex\_lite}, DB: \texttt{kortex\_lite} \\
Automated Enrollment & \texttt{automated} & 1.0 & Web root, PHP/MySQL \\
Tailor Management & \texttt{tailor} & 1.0 & \texttt{/tailor}, DB: \texttt{tailor} \\
Fruits Bazar & \texttt{ecommercefruitsbazarmaster} & 1.0 & Web root, DB: \texttt{ecommerce} \\
Blood Bank System & \texttt{bloodsystem} & 1.0 & Source dir: \texttt{BBfile}, MySQL \\
Covid19 Testing Management System & \texttt{covid19tms} & 1.0 & \texttt{/covid-tms}, DB: \texttt{covidtmsdb} \\
Dairy Farm Management & \texttt{dailyfarmshop} & 1.1 & \texttt{/dfsms}, DB: \texttt{dfsms} \\
CodeAstro Membership Management & \texttt{codeastro} & 1.0 & Web root, DB: \texttt{membershiphp}, PHP 7 recommended \\
Loan Management & \texttt{loansystem} & 1.0 & Source dir: \texttt{Source\_Code}, MySQL \\
Daily Expense Tracker & \texttt{dailyexpensetrackerproject} & 1.0 & \texttt{/dets}, DB: \texttt{detsdb} \\
Visitor Management & \texttt{visitormanagementsystem} & 1.0 & Web root, DB: \texttt{db\_vms} \\
Keerti Login System & \texttt{keerti} & 1.0 & Web root, DB: \texttt{login} \\
\bottomrule
\end{tabular}
\end{table*}

Table~\ref{tab:benchmark-app-versions} reports the dataset identifier, affected
version, and local deployment configuration for each benchmark application. We
selected applications using the following criteria: the source code must be
publicly available, the vulnerable version must be identifiable, the application
must contain at least one reproducible SQL injection or XSS vulnerability, and the
application must be deployable locally with standard PHP/MySQL-style
dependencies.

For each application, we used the source snapshot listed in the table as the
dataset identifier. When an application exposed an internal version number, we
used that value directly. Otherwise, we recovered the affected version from the
public vulnerability record and used the local source-directory name to identify
the evaluated snapshot. Each application was deployed under the listed web root
and initialized with the corresponding database schema or seed file distributed
with the application package.

All experiments use the same source snapshot, database state, base URL, and
filesystem path for a given application. A reported vulnerability is counted only
when it reaches the corresponding vulnerable endpoint and matches the expected
vulnerability class.

\end{document}